\documentstyle[bo99,multicol,epsfig]{article}

\makeatletter
\def\outline{\ifnum \@itemdepth >3 \@toodeep\else
      \advance\@itemdepth \@ne
      \edef\@itemitem{labelitem\romannumeral\the\@itemdepth}\list
      {\csname\@itemitem\endcsname}{\def\makelabel##1{\hss\llap{##1}}
         \parsep \z@ \itemsep \z@
         \ifnum \@enumdepth > 1 \topsep \z@ \fi}\fi}

\makeatother

\def\B{{\em BeppoSAX}}
\def\SM{\mbox{M$_\odot$}}
\def\mcc#1{\multicolumn{1}{|c}{#1}}
\def\be{\begin{equation}}
\def\ee{\end{equation}}
\def\lesssim{\mathrel{\hbox{\rlap{\hbox{\lower4pt\hbox{$\sim$}}}\hbox{$<$}}}}
\def\gtrsim{\mathrel{\hbox{\rlap{\hbox{\lower4pt\hbox{$\sim$}}}\hbox{$>$}}}}

\title{Hard X--ray tails and cyclotron features in X--ray pulsars}

\author{Mauro Orlandini and Daniele Dal Fiume}
\affil{TeSRE Institute, CNR, Bologna, Italy}

\begin{document}

\maketitle

\begin{abstract}

We review the physical processes occurring in the magnetosphere of accreting
X--ray pulsars, with emphasis on those processes that give rise to observable
effects in their high (E$>$10~keV) energy spectra. In the second part we
compare the empirical spectral laws used to fit the observed spectra with
theoretical models, at the light of the BeppoSAX results on the broad-band
characterization of the X--ray pulsar continuum, and the discovery of new
(multiple) cyclotron resonance features.



\end{abstract}

\section{Introduction}

An X--ray pulsar is, by definition, a celestial source showing pulsed emission
when observed in X--rays. The very first pulsating X--ray source, Centaurus
X--3, was discovered in 1971 by the first scientific X--ray satellite {\em
Uhuru} \cite{1437}. Its 4.8~s pulse period implied a small emitting region, and
because the object responsible for the pulsation is not destroyed by the
centrifugal force it is necessary that at its surface the gravitational force
is greater than the centrifugal one. This implies $\Omega_p \la \sqrt{G\rho}$,
where $\Omega_p$ is the pulse frequency, $G$ the gravitational constant, and
$\rho$ the object mean density.  The observed value of $\Omega_p$ implies $\rho
\ga 10^7$ g/cm$^3$ and therefore the compact nature of the object responsible
of the pulsed emission was established.  The binary nature of Cen X--3 was soon
after recognised by the observation of Doppler modulation in the observed pulse
period \cite{648}.  The 2.1~day modulation was coincident with the periodic
disappearing of the source X--ray flux, interpreted as eclipse of the compact
object by the companion. Finally the optical counterpart was discovered as an
early-type O star \cite{655}.  With all these elements it was possible to
determine the mass of the compact object, which resulted to be 1.4~\SM: a
neutron star (NS).

\section{Physical processes in X--ray pulsars}

The physical scenario able to explain the production of pulsed X--ray emission
was elaborated by Shklovskii \cite{845} {\em before} the discovery of Cen X--3.
X--rays are produced in the conversion of the kinetic energy of the accreted
matter (coming from the intense stellar wind of an early-type star --- wind-fed
binaries, or coming from an accretion disc due to Roche-lobe overflow ---
disk-fed binaries) into radiation, because of the interaction with the strong
magnetic field of the NS, of the order of $10^{11}$--$10^{13}$
gauss\footnote{Obtained from conservation of magnetic flux during the process
of collapse from a ``normal'' star ($B\sim 10$--100 gauss, $R\sim 10^6$~Km) to
a NS ($R\sim 10$~Km)}. The dipolar magnetic field of the NS drives the accreted
matter onto the magnetic polar caps, and if the magnetic field axis is not
aligned with the spin axis, the NS acts as a ``lighthouse'', giving rise to
pulsed emission when the beam (or the beams, according to the geometry) crosses
our line of sight.

For a detailed description of the spectral properties of accreting X--ray
pulsars (AXPs) it is therefore necessary to describe the interactions of the
X--rays produced at the NS surface with the highly magnetized plasma forming
the magnetosphere. This is a formidable task because we cannot use a linearized
theory for the radiative transfer equations but we have to deal with the fully
magnetohydrodynamical system. This is due to the fact that the coupling
constants among the interactions are so large that a series expansion is
impossible.

This is the reason why there is not a {\em parametrized} description of AXP
spectra in terms of physical quantities, but only empirical laws to fit
the observed spectra. An alternative method is the numerical solution of the
radiative transport equations assigning particular values to the physical
parameters, comparing the obtained spectra with the observed ones, and varying
the parameters until a match is reached.

\subsection{Cyclotron resonant features}

At some distance from the NS, that we will call magnetospheric radius $r_m$,
the motion of the accreted matter will be dominated by its intense magnetic
field. We define magnetosphere the region around the NS delimited by $r_m$. The
electrons present in the magnetosphere will have an helicoidal motion along the
magnetic field lines, with gyromagnetic (Larmor) frequency given by

\be
\omega_c = \frac{eB}{\gamma mc}   \label{omegac}
\ee

\noindent where $\gamma$ is the Lorentz factor. For the magnetic field strength
$B$ expected in the NS magnetosphere, the motion of the electron in the
direction perpendicular to $B$ is quantized in the so-called Landau levels (see
e.g.\ \cite{1614}). In the nonrelativistic case, the energy
associated to each level is given by

\be
\hbar\omega_n = n\,\hbar\omega_c
\label{omegan}
\ee

\noindent where $\omega_c$ is the Larmor gyrofrequency given in
Eq.~\ref{omegac}. As an aside, from Eq.~\ref{omegan} we have that $E_n =
11.6\cdot B_{12}$~keV, where $B_{12}$ is the magnetic field strength in units
of $10^{12}$ gauss. Therefore we expect to observe cyclotron features in the
hard ($E>10$~keV) energy range.  As we have seen, in the non-relativistic case
the energy levels are harmonically spaced. When relativistic corrections are
taken into account a slight anharmonicity is introduced in the Landau
levels. Indeed, we have

\be
\hbar\omega_n = mc^2\, 
  \frac{\sqrt{mc^2 + 2n\hbar\omega_c\,\sin^2\theta} - 1}{\sin^2\theta}
\ee

\noindent where $\theta$ is the angle between the line of sight and $B$.

Another consequence of the existence of the Landau levels is that an
electromagnetic wave propagating in such a plasma will have well defined
polarization normal modes, {\em i.e.} the medium will be birifringent
\cite{865}.  It is not our intention to enter into the details of the
propagation of waves in the magnetospheric plasma; we will develop a
semi-quantitative approach by highlighting the plasma properties that have
observable consequences in the AXP spectra.

If we introduce the complex refraction index $N$, with its real part the
geometric refraction index and with its imaginary part the absorption
coefficient, then the dispersion relation in the non-relativistic case can be
written as a bi-quadratic equation in $N$. The solution for $N$ will have the
form \cite{1614}

\be
N_1^2 \propto \frac{1}{\omega - \omega_c} \hspace{3cm}
N_2^2 \propto \frac{1}{\omega + \omega_c} \quad.
\ee

The wave with $N_1$ presents resonance and is right-handed circularly polarized
(that is in the same sense as the electron gyration). This wave is called
{\em extraordinary}, in opposition to the {\em ordinary} wave --- described by
$N_2$, which is left-handed circularly polarized. By introducing the
complex refractive index is straightforward to obtain the cyclotron absorption
cross section. By means of the optical theorem we obtain

\be
\sigma_c = 4\pi^2 \alpha_f \frac{\hbar}{m}|e_1|^2\delta(\omega - \omega_c) 
\label{sigmac}
\ee

\noindent where $\alpha_f = e^2/\hbar c$ is the fine structure constant, and 
$\vec{e_1}$ is the polarization versor of the extraordinary wave.

Up to now, we worked neglecting both relativistic corrections and thermal
motions (cold plasma approximation). The release of the latter condition allows
an electron to absorb waves not only of frequency $\omega = \omega_c$, but in
the interval $\omega_c\pm \Delta\omega_D$, where the Doppler width is given by

\be
\Delta\omega_D = \omega_c \sqrt{\frac{2kT}{mc^2}} \, |\cos\theta|
\label{doppler}
\ee

\noindent where $T$ is the electron temperature (we assumed a Maxwell-Boltzmann
distribution for the electrons).

Once the electron absorbs a photon it (almost) immediately de-excitates on a
time scale $t_r\sim 2.6\times 10^{-16}\,B_{12}^{-1}$~sec \cite{1614}. This has
important consequences for the scattering cross sections. Indeed, while a
scattering process involves two photons (one going in, one going out),
absorption (or emission) processes involve only one photon. Therefore one
expect that the two cross section are different. This is not true just because
an absorbed photon is immediately re-emitted, and therefore the
absorption-emission process is equivalent to a scattering. It is possible to
show \cite{1614} that the cyclotron scattering cross section $\sigma_{\rm res}$
has the same form as the cyclotron absorption cross section $\sigma_c$
(Eq.~\ref{sigmac}) with the prescription

\be
\delta(\omega - \omega_c) \rightarrow
\frac{\Gamma_r/2\pi}{(\omega - \omega_c)^2 + \Gamma_r^2/4}
\label{lorenzian}
\ee

\noindent where $\Gamma_r = \gamma_r\omega_c$, and $\gamma_r =
(2/3)(e^2/mc^3)\omega$ is the radiative damping.

Therefore photons with frequency close to  $\omega_c$ will be scattered out of
the line of sight, creating a drop in their number. Cyclotron ``lines''
observed in the spectra of AXPs are therefore {\em not} due to absorption
processes, but are due to scattering of photons resonant with the
magnetospheric electrons (as it occurs for the Fraunhofer lines in the Solar
spectrum). This is why we will not use the term cyclotron lines but the more
appropriate ``cyclotron resonant features'' (CRFs). 

When relativistic effects are taken into account, it is possible to show that
also ordinary waves show resonance, and are scattered out of the line of sight. 
Another important effect on the radiative properties of the plasma is due to a
pure quantum effect: the so-called vacuum polarization. The magnetic field at
which a quantum mechanical treatment of the plasma is necessary can be defined
when the classical cyclotron energy $\hbar\omega_c$ becomes equal to the
electron rest mass $mc^2$. That is

\be
B_{cr} = \frac{m^2c^3}{e\hbar} = 4.414\times 10^{13}\ {\rm gauss} \quad.
\ee

\begin{figure}
\centerline{\psfig{file=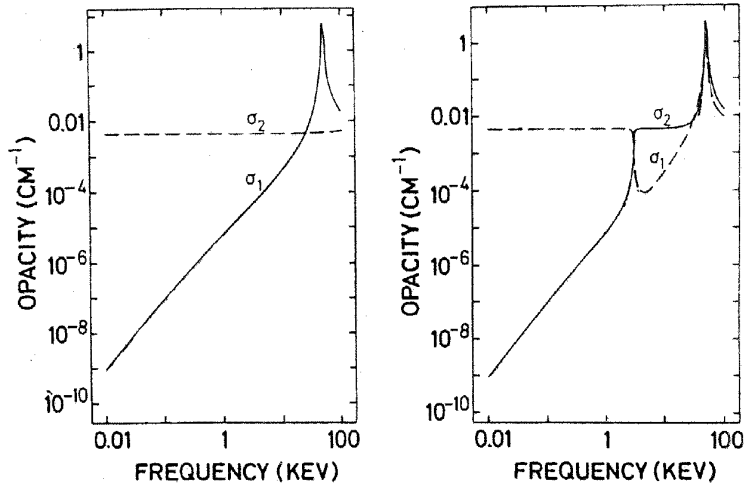,width=0.8\textwidth}}
\caption[]{Angle-averaged scattering opacities (cross section times density)
for the extraordinary ($\sigma_1$) and ordinary ($\sigma_2$) modes, computed
for $n_{22}=1$ and $\hbar\omega_c=50$~keV. {\em Left:} without vacuum
polarization; {\em Right:} with vacuum polarization. From Ventura et~al.\
(1979).}
\label{ventura}
\end{figure}

We call $B_{cr}$ the ``critical'' magnetic field strength. For $B$ not far from
$B_{cr}$ virtual electron-positron pairs can be created. These virtual photons
dominate the polarization properties of the plasma for frequencies in the range
$\omega_{v1} \la \omega \la \omega_{v2}$, where \cite{323,320}

\be
\omega_{v1} \simeq 3\,{\rm keV}\, \frac{\sqrt{n_{22}}}{B/0.1B_{cr}}
\quad ; \hspace{2cm}
\omega_{v2} \simeq \omega_c
\ee

\noindent and therefore affect the scattering cross sections ($n_{22}$ is the
electron density in units of $10^{22}$~cm$^{-3}$). In Fig.~\ref{ventura} we
show the effects of the inclusion of vacuum polarization on the opacity (cross
section times density) as obtained by \cite{918}.

\subsection{Continuum emission}

The main physical process responsible for the continuum emission in AXPs is
Compton scattering.  We will not enter into the details of the problem of
repeated scatterings in a finite, thermal medium (see e.g.\ \cite{868}). Let us
only summarize that an input photon of energy $E_i$ will emerge from a cloud of
non-relativistic electrons (at a temperature $T$) with an average energy
$E_f\sim E_i\,e^y$ (this is valid in the regime $E_f\ll 4kT$). The
comptonization parameter $y$ therefore gives a measure of the photon energy
variation in traversing the plasma, and is given by

\be
y = \left\{ \begin{array}{ll}
  \displaystyle
  \frac{4kT}{mc^2}\, \max\,(\tau, \tau^2) & {\rm Nonrelativistic} \\ \\
  \displaystyle
  \left(\frac{4kT}{mc^2}\right)^2 \, \max\,(\tau, \tau^2) & {\rm Relativistic} \\
 \end{array}   \right.
\ee

\noindent where $\max\,(\tau$, $\tau^2)$ is nothing else but the average
number of scattering suffered by the photons ($\tau$ is the optical depth of
the medium). Note that if $E_i<4kT$ then photons can increase their energy at
the expense of the electrons: this is {\em inverse} Compton scattering.

\begin{figure}
\centerline{\psfig{file=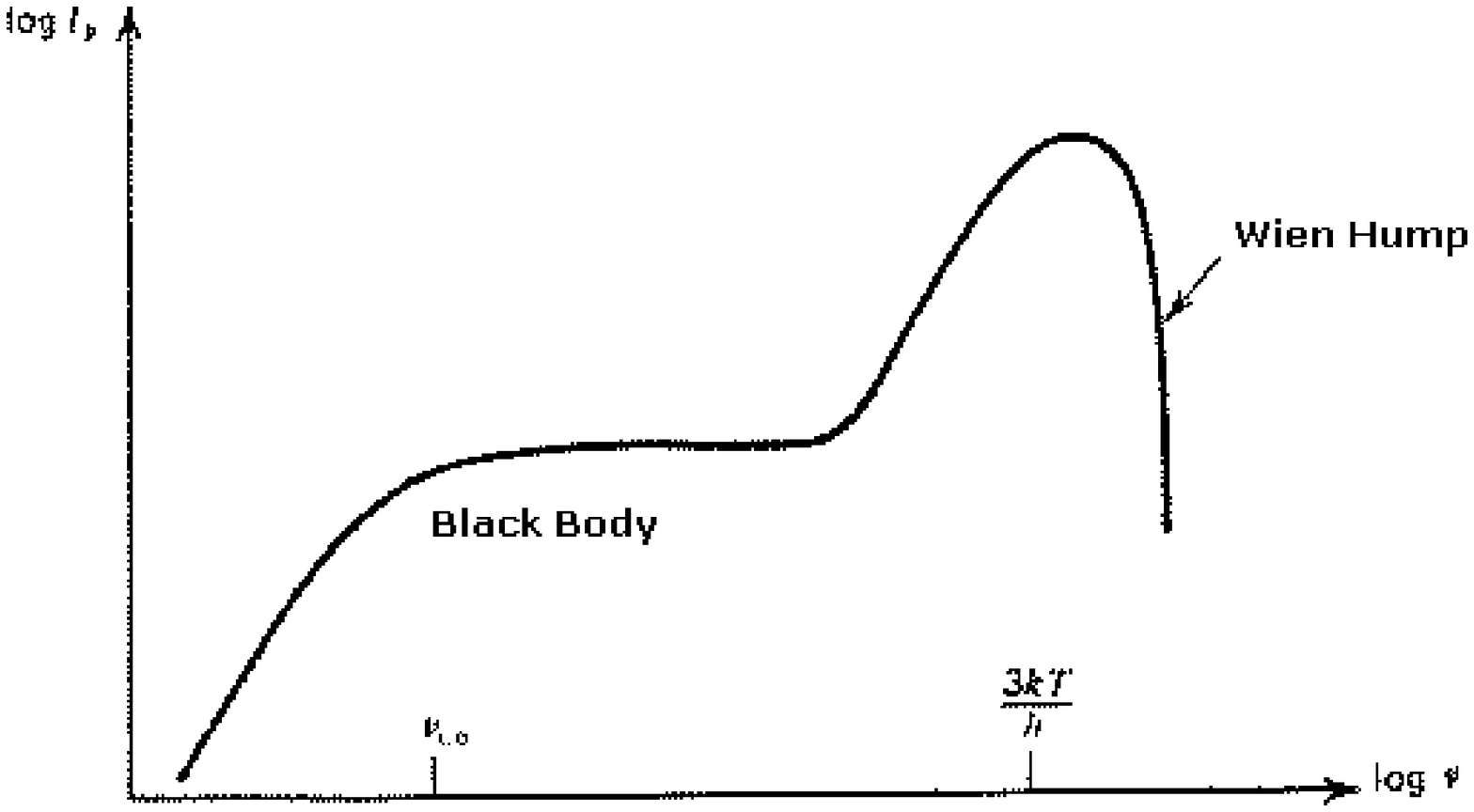,width=0.45\textwidth}%
            \psfig{file=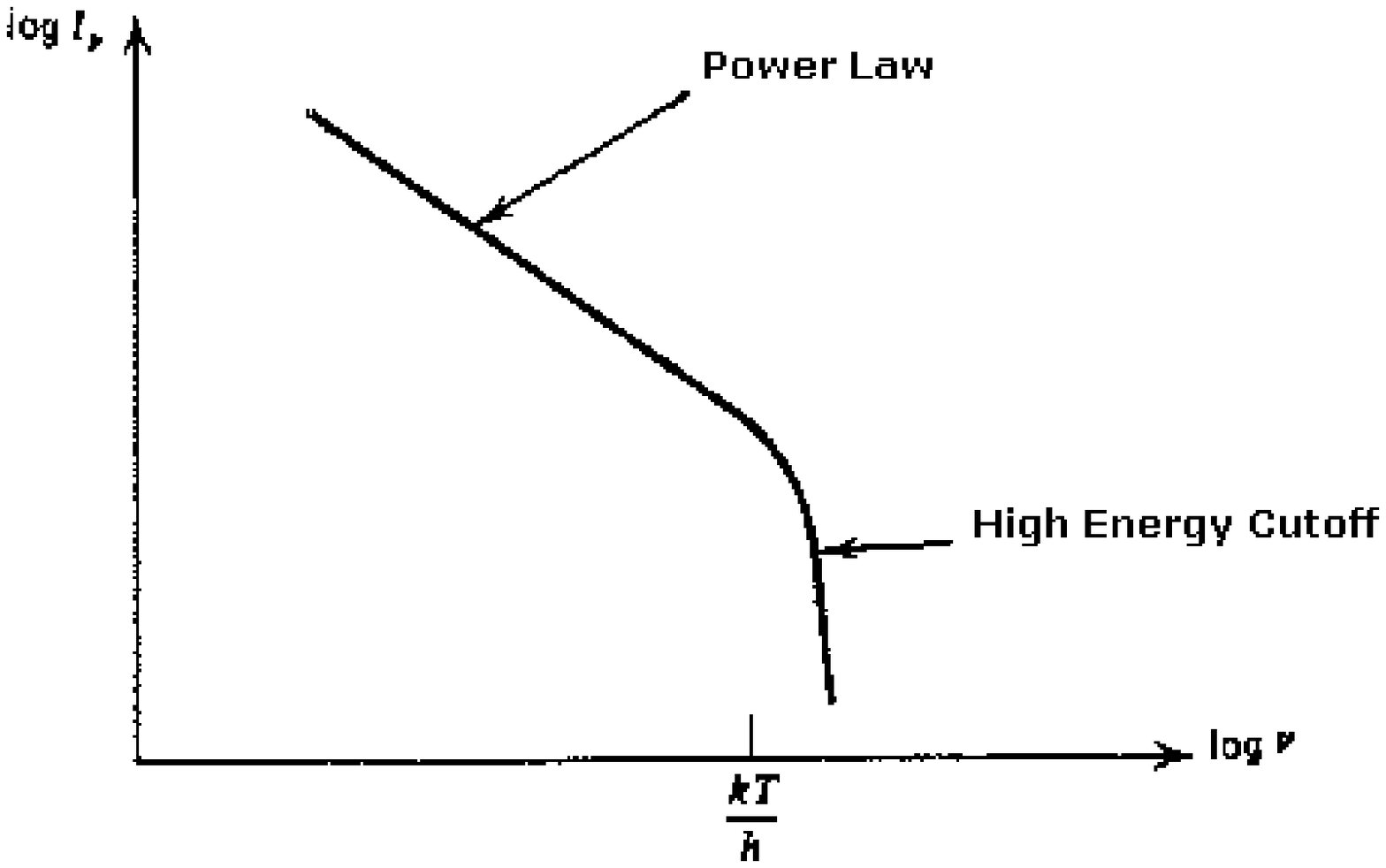,width=0.45\textwidth}}
\caption[]{Spectrum due Compton scattering in a thermal, nonrelativistic
medium. {\em Left:} At low frequency the spectrum is blackbody, while at higher
frequencies develops a Wien hump due to saturated inverse Compton. {\em Right:}
Spectrum produced by unsaturated inverse Compton scattering (adapted from
\cite{868}).}
\label{rybicki}
\end{figure}

The detailed description of the spectrum of the emergent photons requires the
solution of the Kompaneets equation, but it is possible to obtain qualitative
information for special cases:

\begin{outline}
\item $y\ll 1$ In this case only coherent scattering is important, and the
emergent spectrum will be a blackbody spectrum or a ``modified'' blackbody
spectrum according whether the photon frequency is lower or greater than the
frequency at which scattering and absorption coefficients are equal
\cite{868}.

\item $y\gg 1$ Inverse Compton scattering can be important. If we define a
frequency $\omega_{co}$ such that $y(\omega_{co})=1$, then for
$\omega\gg\omega_{co}$ the inverse Compton scattering is saturated and the
emergent spectrum will show a Wien hump, due to low-energy photons up-scattered
up to $\hbar\omega\sim 3kT$ \cite{868}. In the case in which there is not
saturation a detailed analysis of the Kompaneets equation shows that the
spectrum will have the form of a power law modified by a high energy cutoff
\cite{868,614}. These two regimes are qualitatively depicted in
Fig.~\ref{rybicki}.

\end{outline}

\section{Spectral X--ray observations of AXPs}

\subsection{Before \B}

The very first observation of a CRF in a spectrum of an X--ray pulsar was
performed
in 1978 when Tr{\"u}mper et~al.\ \cite{576} observed a $\sim$35~keV CRF in
the spectrum of Hercules X--1. A while later it was observed not only the
fundamental but also the first harmonics in the spectrum of the transient
X--ray pulsar 4U0115+63 \cite{35}. Observations of CRFs in other AXPs showed
that they are a quite common phenomenon in this class of objects. But it was
with the advent of the Japanese satellite Ginga that a systematic analysis of
the spectra of AXPs was performed in search of CRFs. Mihara \cite{1547}
analysed the spectra of 23 AXPs and found that 11 among them showed CRFs.

\subsubsection{Continuum characterization}

Because CRFs are broad features, the exact determination of the continuum is of
paramount importance. From the analysis of the HEAO-1/A2 spectra of AXPs
White et~al.\ \cite{303} found an empirical law that was able to fit
their energy spectra

\be
{\rm POHI}(E) = \left\{ \begin{array}{ll}
 E^{-\alpha}                                           & E < E_{cut} \\
 \displaystyle
 E^{-\alpha} \exp \left(-\frac{E-E_{cut}}{E_f}\right)  & E > E_{cut}
  \end{array}  \right.
\ee

It is evident that this model tries to simulate the unsaturated inverse Compton
process shown in Fig.~\ref{rybicki}. But this model suffers the problem of a
too abrupt break around the cutoff energy $E_{cut}$, problem enhanced by
following more sensitive instruments. Therefore Tanaka \cite{1584} introduced a
``smoother'' cutoff of the form

\be
{\rm FDCO}(E) = \frac{1}{1+\exp{\displaystyle\left(
 \frac{E-E_{cut}}{E_f}\right)}}
\ee

\noindent that he called Fermi-Dirac cutoff because of it resemblance with the
Fermi-Dirac distribution function. It is important to stress that the FDCO
model does not have any physical meaning: it only gives a better description of
the break in the AXP spectra.

Makishima and Mihara \cite{407} were the first to note that in the AXPs showing
CRFs there was a correlation between the cutoff energy $E_{cut}$ and the CRF
energy $E_c$, namely $E_c\simeq (1.2-2.5)\cdot E_{cut}$. Therefore it
seemed that the cutoff was in some way due to the presence of the CRF.  The
next step was performed by Mihara, who introduced the so-called NPEX (Negative
Positive EXponential) model

\be
{\rm NPEX}(E) = (AE^{-\alpha} + BE^{+\beta})
  \exp\left(-\frac{E}{kT}\right) \quad.
\ee

This model is quite successful in describing the AXP spectra observed by Ginga
in the 3--30~keV. Its components have also a physical meaning, because it
mimics the saturated inverse Compton spectrum shown in Fig.~\ref{rybicki} if
$\beta=2$. Furthermore, because the (non relativistic) energy variation of a
photon during Compton scattering is \cite{868}

\be
\frac{\Delta E}{E} = \frac{4kT-E}{mc^2}
\ee

\noindent then when $E=E_c$ the medium is optically thick and therefore
$E_c\sim 4kT$.

\subsubsection{CRF characterization}

Mihara, besides the introduction of the NPEX model to describe the AXP
continuum, introduced a new form for the CRF

\be
{\rm CYAB}(E) = \exp \left(-\frac{\tau(WE/E_c)^2}{(E-E_c)^2 + W^2} \right)
\ee

\noindent which has the form of a Lorenzian of width $W$, and depth $\tau$.
From a physical point of view, this is the form assumed by the  cyclotron
scattering cross section described by Eq.~\ref{lorenzian}.

\subsection{\B\ observations of AXPs}

With the advent of \B\ the study of energy spectra of AXPs received new impulse
because it was now possible to characterize with unprecedent detail the
continuum on a broader energy range (0.1--200~keV), and the two high energy
instruments aboard \B, namely HPGSPC (sensitive in 5--60~keV; \cite{1533}) and
PDS (15-200~keV; \cite{1386}), are the best suited for the detailed
spectroscopy of CRFs. \B\ observed all the persistent AXPs, plus a couple of
transient ones (see Table~\ref{table}). As a first result, we found that the
NPEX model, successfully used to fit the Ginga data, is not adequate to
describe the broad AXP continuum \cite{1946}. In particular we find that their
continuum can be described in terms of (i) a black-body component with
temperature of few hundreds eV; (ii) a power law of photon index $\sim$1 up to
$\sim$10~keV; and a (iii) a high energy ($\ga$10~keV) cutoff that makes the
spectrum rapidly drop above $\sim$40--50~keV.

\begin{table}
\begin{center} 
\begin{tabular}{|l|l|c|c|l|} 
\hline\hline\noalign{\smallskip}
\mcc{Source}& \mcc{Obs Date}  & \mcc{E$_{\rm cyc}$ (keV)} & \mcc{FWHM (keV)} & \multicolumn{1}{|c|}{References} \\ 
\hline\hline\noalign{\smallskip} 
{\bf 4U0115+63 (M)}    & 20 Mar 1999 & $12.78\pm 0.08$ & $3.58\pm 0.33$ & \cite{1933} \\ 
{\bf 4U1538--52 (M)}   & 29 Jul 1998 & $21.5\pm 0.4$   & $6.7\pm 1.2$   & \cite{robba99} \\ 
{\bf Cen X--3 (M?)}    & 27 Feb 1997 & $28.5\pm 0.5$   & $7.3\pm 1.9$   & \cite{1802} \\ 
{\bf XTE J1946+27}     & 09 Oct 1998 & $33\pm 4$       & $16\pm 2$      & \cite{segreto99} \\ 
{\bf OAO1657--415}     & 04 Sep 1998 & $36\pm 2$       & 10             & \cite{1961} \\ 
{\bf 4U1626--67}       & 06 Aug 1996 & $38.0\pm 0.9$   & $11.8\pm 1.7$  & \cite{1622} \\ 
{\bf 4U1907+09 (M)}    & 29 Sep 1997 & $38.3\pm 0.7$   & $9.7\pm 2.3$   & \cite{1736} \\ 
{\bf Her X--1}         & 27 Jul 1996 & $42.1\pm 0.3$   & $14.7\pm 1.1$  & \cite{1583} \\ 
{\bf GX301--2 (M)}     & 24 Jan 1998 & $49.5\pm 1.0$   & $17.9\pm 2.5$  & \cite{1803} \\ 
{\bf Vela X--1 (M)}    & 14 Jul 1996 & $54.8\pm 0.9$   & $25.0\pm 2.1$  & \cite{1581} \\ 
{\bf GX1+4}            & 25 Mar 1997 & \ldots          & \ldots         & \cite{1652} \\ 
{\bf GS1843+00}        & 04 Apr 1997 & \ldots          & \ldots         & \cite{1741} \\ 
{\bf X Persei}         & 09 Sep 1996 & \ldots          & \ldots         & \cite{1810} \\ 
\hline\noalign{\smallskip} \multicolumn{5}{l}{M stands for multiple lines detected/suspected} 
\end{tabular} 
\end{center}
\caption[]{BeppoSAX observations of X--ray binary pulsars} \label{table}
\end{table}

Furthermore, the CRFs observed with \B\ are better described in terms a
Gaussian in absorption, defined as \cite{1172}

\be
{\rm GAUABS}(E) = \left[1-I\exp\left(-\frac{(E-E_c)^2}{2W^2}\right)\right]
\quad.
\ee

In order to better characterize the CRF we introduced a new tool, the so-called
normalized Crab ratio. The Crab ratio is simply the ratio between the source
count rate spectrum and the count rate spectrum of the Crab Nebula. As this
second spectrum is known, with great accuracy, to be free of features and to be
modeled at first order with a power law in a very broad energy range, this
ratio is quite well suited to enhance the presence of features in the spectrum.
Furthermore the ratio is in first approximation independent from the
calibration of the instrument.

\begin{figure}
\centerline{\psfig{file=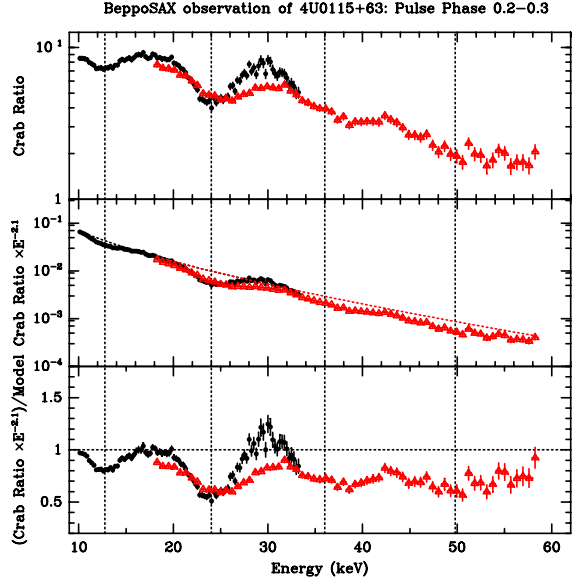,width=0.7\textwidth}}
\vspace{-1cm}
\caption[]{Normalized Crab ratio for the \B\ observation of the transient AXP
4U0115+63. Both the two high energy instruments, HPGSPC (black marks) and PDS
(grey marks), are shown. Note the presence of at least four cyclotron harmonics
\cite{1933}.}
\label{crab-ratio}
\end{figure}

In order to enhance the deviations from the continuum we multiply the ratio by
a E$^{-2.1}$ power law, that is the functional form of the Crab Nebula
spectrum, and we divide by the functional describing the continuum shape of the
source (from this the name normalized Crab ratio). The procedure is described
in Fig.~\ref{crab-ratio} where we plot the result of each different step used
to obtain the final result in the case of 4U0115+63. Note the presence of up to
four cyclotron harmonics in the spectrum \cite{1933}.

\begin{figure}
\centerline{\psfig{file=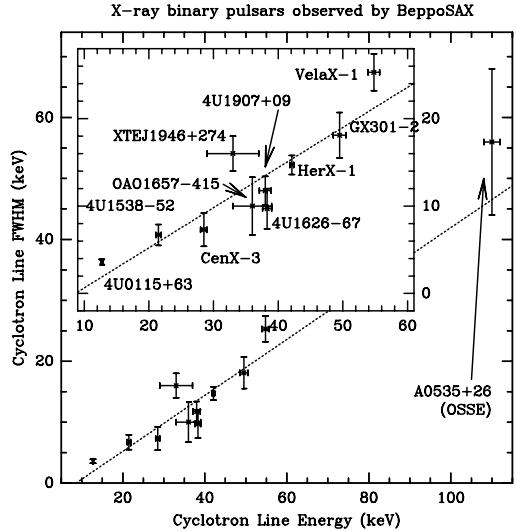,height=8cm}}
\vspace{-1cm}
\caption[]{FWHM vs centroid energy for the cyclotron features observed
by \B\ and listed in Table~\ref{table}. We include the OSSE measurement on the
1994 outburst of A0535+26 \cite{375}. Note that the linear correlation has
been computed {\em without} taking into account the A0535+26 point.}
\label{CRF_vs_FWHM}
\end{figure}

By performing the normalized Crab ratio on all the AXPs listed in
Table~\ref{table} it is immediate to observe that higher the CRF energy,
broader the feature \cite{1946}. The correlation between $E_c$ and the CRF FWHM
is quite evident in Fig.~\ref{CRF_vs_FWHM}, and is easily understood in terms
of Doppler broadening of the electrons responsible of the resonance, and holds
for all the sources displaying single CRFs (see Eq.~\ref{doppler})\footnote{We
assume that CRFs are produced quite close to the NS surface, therefore
neglecting the gravitational redshift, which shifts the centroid energy by a
factor $(1+z)^{-1} = (1-2GM/Rc^2)$. See discussion in \cite{1946}.}.  It is
important to stress that this relation does not hold in presence of multiple
harmonics. In other words, it seems that the temperature of the electrons
responsible of higher CRF harmonics is different from that of the electrons
responsible of the fundamental CRF. From Fig.~\ref{CRF_vs_FWHM} and by means of
Eq.~\ref{doppler} we derived that the electron temperature responsible for the
resonance is in the range $\sim$15--30 keV. This energy range is somehow
``critical'', because in some AXPs (and the effect is particularly evident in
OAO1657--40 \cite{1961}) we observe actually {\em two} changes of slope in the
high energy part of their spectra: a first change of slope occurs in the
$\sim$10--20~keV range, while a second steepening occurs for higher energies.
This leds to our last issue: ``anomalous'' multiple CRFs.

There are three sources, namely 4U1907+09, Vela X--1, and GX301--2 that
require two CRFs in their energy spectra. The anomaly is that (i) the two CRFs
are not harmonically spaced; (ii) the depth of the ``fundamental'' is much
smaller than that of its ``harmonics'', (iii) their width does not correlate
with their centroid energy and, more importantly, (iv) there is no trace of the
``fundamental'' in the normalized Crab ratio. Because of this last point we
have some doubt about the interpretation of them as CRF, expecially because
they are all in the critical energy range 10--30~keV where we observe the
change of slope in the continuum. We are therefore inclined to interpret them
as due to a not correct modelization of the continuum. Another possible
explanation could be that they are due to vacuum polarization effects (see
Fig.~\ref{ventura}), but this interpretation requires a more quantitative
analysis. For the three sources discussed above, we did not plot in
Fig.~\ref{CRF_vs_FWHM} the ``anomalous'' CRF but the one obtained from the
normalized Crab ratio.

\section{Conclusions}

The broad-band capabilities of \B\ have shown that the simple phenomenological
spectral laws used to describe AXP spectra in narrow energy ranges are
inadequate to fit broad-band spectra. The study of AXPs as a class has shown
that there is a critical region between $\sim$10 and $\sim$30~keV in which we
observe a change of slope in the continuum. If not well modeled this could give
rise to extraneous features that could be interpreted as CRF. Probably a
detailed treatment of Compton scattering taking into account the effects of the
magnetic field could help to solve this issue. Also vacuum polarization effects
could alter the emergent energy spectra and explain the observed ``anomalous''
CRF harmonics.

Doppler broadening of the electrons responsible for the CRF is able to explain
the observed correlation between CRF FWHM and centroid energy, showing that the
electron temperature is in the range 15--30~keV for all the observed AXPs. This
correlation does not hold for sources showing multiple CRFs, implying that the
temperature of the electrons giving rise to higher harmonics could be
different.

\medskip

\noindent {\em Acknowledgements} We wish to thank the ``X--ray pulsar fans''
working group, formed by the friends at the TeSRE, IFCAI and ESTEC/SSD
institutes in Bologna, Palermo and Noordwijk, who produced a good wealth of
results on \B\ observations and without whom this work would not have been
possible.

\bibliographystyle{apj}
\begin{multicols}{2}
\small\bibliography{mauro:[bibtex.bib]bib,mauro:[tex]bib1}
\end{multicols}

\end{document}